\documentstyle[prl,aps,tighten,multicol]{revtex}

\begin{document}


\title{ On the scaling laws for the largest Lyapunov
exponent \\ in long-range systems: A random matrix approach }

\author{Celia Anteneodo\thanks{{\rm e-mail: celia@cbpf.br}}
and
Ra\'ul O. Vallejos\thanks{{\rm e-mail: vallejos@cbpf.br}}
}

\address{Centro Brasileiro de Pesquisas F\'{\i}sicas,
R. Dr. Xavier Sigaud 150, \\
22290-180, Rio de Janeiro, Brazil}


\maketitle

\begin{abstract}
We investigate the laws that rule the behavior of the
largest Lyapunov exponent (LLE) in many particle systems with long range
interactions.
We consider as a representative system the so-called Hamiltonian
$\alpha$-XY model where the adjustable parameter $\alpha$ controls
the range of the interactions of $N$
ferromagnetic spins in a lattice of dimension $d$.
In previous work the dependence of the LLE with the system size $N$, for
sufficiently high energies, was established through numerical simulations.
In the thermodynamic limit, the LLE becomes constant for $\alpha>d$ whereas
it decays as an inverse power law of $N$ for $\alpha<d$.
A recent theoretical calculation based on a geometrization of the dynamics
is consistent with these numerical results.
Here we show that the scaling behavior can also be explained by a
random matrix approach, in which the tangent mappings that define the
Lyapunov exponents are modeled by random simplectic matrices drawn from a
suitable ensemble. \\[3mm]
\end{abstract}


\begin{multicols}{2}

\narrowtext


\section{Introduction}


Dynamical systems of many particles interacting via long range forces can
exhibit interesting anomalies such as
super-diffusion \cite{ator98,tora99,lrr99},
metastable states \cite{lrr00} and
non Boltzmann-Gibbs distribution functions \cite{lrt00}.
Special interest in such systems has arisen recently in
connection with the non-extensive generalization of statistical mechanics
introduced by Tsallis \cite{ct01}.

The existence of a dynamics makes the systems mentioned above
very attractive because it is possible, in principle,
to associate the properties of the thermodynamic states with
features of the many particle phase space. As a remarkable example, 
let us mention the ``Topological Hypothesis", which relates thermodynamic
phase transitions to topological changes in the structure of phase space
\cite{ccp99}.

A dynamical model with an adjustable interaction range $\alpha$, and
allowing extensive numerical and analytical exploration has 
been recently introduced \cite{at98,gmc00,cgmt00}.
The model consists in a periodical $d$-dimensional lattice of $N$
interacting rotators moving on parallel planes. Each rotator
is restricted to the unit circle and therefore it is fully described by
an angle $0 < \theta_i \le 2\pi$ and
its conjugate momentum $L_i$, with $i=1,\ldots,N$.
The dynamics of the system is governed by the Hamiltonian
\begin{equation}
\label{ham0}
H = \frac{1}{2} \sum_{i=1}^{N} L_{i}^{2} \,
+ \frac{J}{2} \sum_{i,j=1, i\neq j }^N
\frac{1 - \cos(\theta_{i} - \theta_{j} )}{r_{ij}^{\alpha} },
\end{equation}
where the coupling constant is $J\geq 0$ and, without loss of generality,
unitary moments of inertia are chosen for all the particles.
Here $r_{ij}$ measures the minimal distance between the rotators
located at the lattice sites $i$ and $j$.
The Hamiltonian (\ref{ham0}) describes a classical
inertial XY ferromagnet.
It contains as particular cases the
mean-field version ($\alpha/d = 0$) and the
first-neighbors case, recovered in the $\alpha/d \to \infty$ limit.

The systems $\alpha$-XY characterized by interaction ranges
$0 \le \alpha/d < 1$ do not have
a well defined thermodynamic limit, e.g.,
the specific (per particle) energy diverges when $N\to\infty$.
A proper thermodynamic limit is defined by introducing a scaling
parameter $\widetilde{N}$ \cite{ct95},
which depends on $N$, $\alpha$, and $d$ \cite{ta00,cgm00}:
\begin{equation}
\widetilde{N} =
\frac{1}{N}
\sum_{i,j=1, i\neq j }^N
\frac{1}{r_{ij}^{\alpha} } \; .
\end{equation}
In the large $N$ limit one has
\begin{equation}
\widetilde{N} (\alpha/d) \sim
\left\{ 
\matrix{ 
  N^{1-\alpha/d} & & 0 \le \alpha < d \cr 
          \ln{N} & & \alpha = d \cr 
\Theta(\alpha/d) & & \alpha > d } \right.  
\label{asym} 
\end{equation} 
with $\Theta$ a function of the ratio $\alpha/d$ only. 
Specific energy-like quantities must be rescaled by $\widetilde{N}$. 
At the dynamic level, time, hence inverse Lyapunov exponents, have to 
be scaled by $\widetilde{N}^{-1/2}$ \cite{at98}. 

A completely equivalent description is obtained by 
working with the already
scaled Hamiltonian
\begin{equation}
\label{ham}
\widetilde{H} = \frac{1}{2} \sum_{i=1}^{N} L_{i}^{2} \,
+ \frac{J}{2 \widetilde{N}}
\sum_{ i,j=1 , i\neq j }^N
\frac{1-\cos(\theta_{i}-\theta_{j})}{r_{ij}^{\alpha} }.
\end{equation}
This kind of scaling of the strength of the interactions, common in
standard mean-field discussions, has been
applied to the study of the mean-field case of the present model in Refs.
\cite{ar95,lrr98}.
Since the Hamiltonian (\ref{ham}) leads to the same results as
(\ref{ham0}), but avoiding further rescalings,
all our considerations from here on will be related to the already scaled
Hamiltonian (\ref{ham}).

The $\alpha$-XY ferromagnet has been subject of several numerical and
analytical studies.
References \cite{at98,ta00,cgm00} and \cite{gmc00,cgmt00} are
dedicated to the cases $d=1$ and $d=2,3$, respectively.
The mean-field problem is discussed in Refs. \cite{ar95,lrr98} while
the opposite limit of first-neighbor interactions
and $d=1$ can be found in  \cite{eklr94}.
For long-ranged interactions, $\alpha/d<1$, the system displays a
second order phase transition from a ferromagnetic state to a paramagnetic
one at a certain critical specific energy $\varepsilon_c$
($\varepsilon_c=0.75 J$) \cite{ta00,cgm00}.
It is likely that systems with $1<\alpha/d<2$ also undergo a second order
phase transition but the critical energy may depend on $\alpha/d$,
as has been shown numerically for $d=1$ \cite{ta00}.
For $\alpha/d>2$, all systems behave similarly to the first-neighbor model,
where there is no order nor phase transition for finite energies in the
thermodynamic limit\cite{gmc00,ta00}.

Here we are concerned with the high energy phase, i.e.,
with energies above $\varepsilon_c$.
In this disordered regime, the kinetic energy is much larger than the
bounded potential energy, the rotators are weakly coupled and
the dynamics is weakly chaotic.
In Refs.~\cite{at98,gmc00,cgmt00} largest Lyapunov exponents
(LLE) were calculated numerically.
It was found that, in the thermodynamic limit, and for large enough
energies, the largest Lyapunov exponent remains positive and finite
for short-range interactions ($ \alpha/d > 1 $) but vanishes as an inverse 
power law of the system size in the long-range case $0 \le \alpha/d < 1$.

Recently Pettini and co-workers developed a theoretical method which,
in principle, allows to obtain the scaling behavior of the LLE analytically
(see \cite{cpc00} for a review).
In this approach the phase space trajectories are mapped onto a geodesic
flow in configuration space (equipped with a suitable metric).
It is assumed that the curvature along an ergodic geodesic can be modeled as
a Gaussian process. Then the LLE is expressed in terms of the
mean and  variance of the process. These parameters are
calculated as microcanonical averages of suitable dynamical functions.
There are several works where the method was applied to the $\alpha$-XY
model.
The scaling behavior of the LLE in the extreme cases $\alpha\to\infty$ and
$\alpha=0$ was found in Refs.~\cite{ccp96} and \cite{f98}, respectively.
Very recently, Firpo and Ruffo\cite{fr01} succeeded in calculating the LLE
scaling laws for any interaction range $ 0 \le \alpha/d < 1$.

It is our purpose here to present a simple alternative procedure,
based on a random matrix formulation,
which allows to derive the dependence of the LLE
on the size of the system $N$, the range
of the interactions $\alpha$ and the lattice dimension $d$.
This procedure is based on the ideas introduced by Benettin \cite{b84}
in the discussion of two dimensional billiards, and later extended to
interacting many particle systems \cite{palv86,parv86}.


\section{The largest Lyapunov exponent}


The equations of motion generated by the Hamiltonian (\ref{ham}) are
\begin{eqnarray}
\nonumber
\dot{\theta_i} & = & L_i \nonumber \\
\dot{L_i} & = & -\frac{J}{\widetilde{N}}\sum_{j=1, j\neq i}^N
\sin{(\theta_i - \theta_j)} /r_{ij}^{\alpha} \; ,
\end{eqnarray}
for $i=1,\ldots,N$.
Discretizing the time axis into steps $\Delta t$ one obtains the
stroboscopic map relating angles and momenta at successive discrete times
\begin{eqnarray}
\nonumber
\theta'_i & = & \theta_i \,+\, L_i \,\Delta t\\ \label{stroboscopic}
L'_i & = & L_i \,-\, \frac{J }{\widetilde{N}}
\sum_{j=1, j\neq i }^N
\sin{(\theta'_i - \theta'_j)} /r_{ij}^{\alpha} \,\Delta t\,.
\end{eqnarray}
For the purpose of discussing Lyapunov exponents one has to consider the
tangent
map $T$, i.e., the linearized version of Eqs.~(\ref{stroboscopic}), 
given by
\begin{eqnarray}
\nonumber
\delta\theta'_i & = & \delta\theta_i \,+\, \delta L_i \,\Delta t \\
\delta L'_i & = & \delta L_i \,-\, \frac{J }{\widetilde{N}}
\sum_{ j=1, j \neq i }^N
\cos{(\theta'_i - \theta'_j)} \,[\delta \theta'_i \,-\, \delta
\theta'_j]/r_{ij}^{\alpha} \,\Delta t.
\end{eqnarray}
In matrix form these equations read
\begin{equation}
\left( \matrix{ \delta\mbox{\boldmath $\theta'$} \cr
\delta\mbox{\boldmath $L' $} }\right) =
\left( \matrix{ \openone & \openone\Delta t \cr
\epsilon\hat{b}\Delta t & \openone + \epsilon \hat{b}
(\Delta t)^2 }\right)
\left( \matrix{ \delta\mbox{\boldmath $\theta$} \cr
\delta\mbox{\boldmath $L$ } }\right) \equiv T
\left( \matrix{ \delta\mbox{\boldmath $\theta$} \cr
\delta\mbox{\boldmath $L$ } }\right) \; ,
\end{equation}
where all submatrices are of size $N\times N$, $\openone$ being the
identity. The matrix $\hat{b}$ is given by
\begin{equation}
\label{hatb}
b_{ij} =
\left\{ 
\matrix{ \cos{(\theta'_i-\theta'_j)}/r_{ij}^{\alpha} 
& \mbox{for} & j \neq i \cr 
\displaystyle{-\sum_{ k=1, k\neq i }^N} 
\cos{(\theta'_i-\theta'_k)}/r_{ik}^{\alpha} 
& \mbox{for} & j=i } 
\right. \;   
\end{equation}
and the perturbation parameter $\epsilon$ is
\begin{equation}
\label{epsilon}
\epsilon=J/\,\widetilde{N}(\alpha/d) \, .
\end{equation}
For long-range interactions it is clear that $\epsilon$ goes to zero in
the thermodynamic limit. In the short-range case one has $\epsilon \sim J$.
In order to treat $\epsilon$ as a small parameter in both cases,
we will consider the limit $J \to 0$ when necessary.

The LLE can be defined by the limiting procedure
\begin{equation}
\label{lyapdef}
\lambda_{max}\,=\, \lim_{n \rightarrow \infty} \frac{1}{n \Delta t}
\ln \left|\left| {\cal T} \xi \right|\right|\; ,
\end{equation}
with $\xi$ an arbitrary vector and
${\cal T} \equiv T_n T_{n-1}\ldots T_2 T_1$
is the product of $n$ tangent maps calculated at successive points of the
discretized trajectory. Using the euclidean norm, Eq.~(\ref{lyapdef}) is
rewritten as
\begin{equation}
\lambda_{max} = \lim_{n \rightarrow \infty} \frac{1}{2n\Delta t}
\ln \left( \xi^t{\cal T}^t{\cal T}\xi \right) \; ,
\label{lle}
\end{equation}
the superscript $t$ indicating ``transposed''.


\section{Random matrix approach}


The random matrix approach is based in the modeling of the
tangent mappings $T_k$ by a sequence of non-correlated random simplectic
matrices mimicking the essential properties of the chaotic dynamics.
In the standard procedure \cite{palv86} one replaces
the short-time $T_k$ by finite time random matrices $R_k$ having the
same structure
\begin{equation}
R_k=
\left( \matrix{ 
\openone & \openone\tau \cr 
\epsilon\hat{a}_k\tau & \openone + \epsilon \hat{a}_k \tau^2}\right) \; . 
\end{equation} 
In previous treatments the time scale $\tau$ has always been ignored by setting 
it to one. Here we prefer to keep track of $\tau$, as later we will argue that it
is related to energy. It is fixed as follows.
The time scale $\tau$ must be chosen small enough in order to preserve 
the short time structure of the tangent maps.
However, it cannot be too small, as we assume that consecutive 
tangent maps are
statistically independent. Thus $\tau$ must be an intermediate scale, of the
order of
the correlation time for the tangent maps.
(This time scale is analogous to the correlation
time of the Gaussian process modeling the fluctuations of the curvature in
the geometric
method \cite{cpc00}.)

The symmetric matrices $\hat{a}_k$ are the random analogs of $\hat{b}_k$ and
are assumed to be independent, i.e.,
$\langle \hat{a}_n \hat{a}_m \rangle=\langle \hat{a}_n\rangle
\langle\hat{a}_m\rangle$,
unless $n=m$. Except for the symmetry restriction, the elements ${a}_{ij}$
of a
given matrix $\hat{a}_k$ are independent \cite{livi}.

The probability distributions of the elements $a_{ij}$ are obtained from
$b_{ij}$ (\ref{hatb}) by considering that the angles $\theta'_i$ are
independent
from each other and uniformly distributed in $[0,2\pi)$.
This assumption, reasonable in the high energy phase, implies that:
(i) The average of each ${a}_{ij}$ is null.
(ii) The information about the range of the
interactions is embodied in the variance of each ${a}_{ij}$, which
depends on the distance between sites:
\begin{eqnarray}
\nonumber
\langle ({a}_{ij})^2 \rangle & =
& \frac{1}{2}\, r_{ij}^{-2\alpha} \qquad (i\neq j) \\
\langle ({a}_{ii})^2 \rangle & =
& \frac{1}{2}\,\widetilde{N}(2\alpha/d) \; .
\end{eqnarray}
In this way one has defined a crude although non trivial statistical
model whose validity has been shown in previous works
\cite{b84,palv86,parv86}.

As a consequence of the assumptions made one has the property
$ \langle \hat{a}_k \hat{a}^t_k \rangle = \gamma \openone $,
which will be useful for evaluating averages. In our case
\begin{equation}
\label{gamma}
\gamma=\widetilde{N}(2\alpha/d) \; .
\end{equation}
Within this model the expression for the LLE is obtained by averaging over
different realizations of sequences of random matrices
\begin{equation}
\lambda_{max} = \lim_{n\rightarrow\infty}
\frac{1}{2n\tau} \langle \ln
\xi^t
R_1^t
\ldots
R^t_{n-1}
R^t_n
R_n
R_{n-1}
\ldots
R_1
\xi \rangle \;.
\end{equation}
Assuming that the distribution of LLE's over the ensemble of sequences
is narrow, one can interchange the average and the logarithm. Then the
averaging scheme is reduced to a sequence of averages over each matrix
distribution
\begin{equation}
\lambda_{max} \simeq \lim_{n\rightarrow\infty}
\frac{1}{2n\tau} \ln
( \xi^t
\langle
\ldots
\langle R^t_{n-1}
\langle
R^t_n
R_n \rangle
R_{n-1} \rangle
\ldots
\rangle
\xi ) .
\end{equation}
These averages have already been calculated by Parisi and Vulpiani
\cite{parv86} (see also \cite{lima90}).
Instead of just recalling their results, we prefer to exhibit a different
derivation, which not only makes the paper self-contained, but may be
interesting by itself. Notice that the first average can be immediately
done,
the result being the symmetric matrix
\begin{equation}
\label{avrr}
\langle R^t_n R_n \rangle =
\left( \matrix{ 
\nu_1 \openone & \sigma_1 \openone \cr 
\sigma_1 \openone & \mu_1 \openone }\right) \, , 
\end{equation} 
with 
\begin{eqnarray} 
\nu_1 & = & 1+\gamma\epsilon^2\tau^2 \nonumber \\ 
\mu_1 & = & 1+\tau^2+\gamma\epsilon^2\tau^4 \nonumber \\ 
\sigma_1 & = & \tau+\gamma\epsilon^2\tau^3 \; . 
\end{eqnarray} 
The remaining $n-1$ averages can be done iteratively, obtaining at each step 
a symmetric matrix with the same structure as (\ref{avrr}). The final result is 
\begin{equation} 
\langle R^t_1 \langle \ldots \rangle R_1 \rangle = 
\left( \matrix{ 
\nu_n \openone & \sigma_n \openone \cr 
\sigma_n \openone & \mu_n \openone }\right) \, , 
\end{equation} 
where the coefficients $\nu, \mu, \sigma$ are calculated by means of the 
recurrence relation 
\begin{equation} 
\left(
 \matrix{\nu_n \cr
\mu_n \cr
\sigma_n }\right)
=
\left( \matrix{ 1 &\gamma\epsilon^2\tau^2 & 0 \cr
\tau^2 & 1+\gamma\epsilon^2\tau^4 & 2 \cr
\tau &\gamma\epsilon^2\tau^3 & 1 }\right)^{n-1}
\left( \matrix{\nu_1 \cr
\mu_1 \cr
\sigma_1 }\right) \; .
\end{equation}
Then it can be easily checked that the LLE is related
to the maximum eigenvalue $L_{max}$ of the $3\times 3$
matrix above through the formula
\begin{equation}
\lambda_{max} = \frac{1}{2\tau} \ln L_{max} \; .
\end{equation}
After solving the cubic eigenvalue equation we expand $L_{max}$ around
$\epsilon=0$,
\begin{equation}
L_{max} \,=\, 1\,+\, (2\gamma\epsilon^2\tau^4)^{1/3} + \cdots \; ,
\end{equation}
so that
\begin{equation}
\lambda_{max} = \frac{1}{2} (2\gamma\epsilon^2\tau)^{1/3} + \cdots \; .
\end{equation}
Finally, substituting $\epsilon$ and $\gamma$ by their definitions
(\ref{epsilon}) and (\ref{gamma}), respectively, one gets the compact
expression
\begin{equation}
\lambda_{max} \propto
J^{2/3}\,\tau^{1/3}\,\left[ \frac{\widetilde{N}(2\alpha/d)}
{\widetilde{N}^2( \alpha/d)} \right]^{1/3} \; .
\end{equation}
However, for the purpose of comparison with previous works, it
is convenient to explicit the $N$ dependence.
Recalling the asymptotic expression (\ref{asym})
for $\widetilde{N}$, we arrive at
\begin{equation}
\label{finalresult}
\lambda_{max} \propto J^{2/3}\,\tau^{1/3}\,
\left\{ 
\matrix{ 
1/N^{1/3} & & 0 \le \alpha/d < 1/2 \cr 
(\ln{N}/N)^{1/3} & & \alpha/d = 1/2 \cr 
1/N^{2(1-\alpha/d)/3} & & 1/2 < \alpha/d <1 \cr 
1/(\ln{N})^{2/3} & & \alpha/d = 1 \cr 
constant & & 1 < \alpha/d 
} \right. \, .
\end{equation} 
This scaling law for the LLE can also 
be written as $\lambda_{max}\sim 1/N^\kappa$ with
\begin{equation}  
\kappa \,=\,  
\left\{ 
\matrix{ 
1/3             & & 0 \le \alpha/d \leq 1/2 \cr 
2(1-\alpha/d)/3 & & 1/2 < \alpha/d <1 \cr 
0               & &  1 < \alpha/d 
} \right. ,
\label{kappa}
\end{equation}
the case $\alpha/d=1$ being marginal.

We expect the scaling above not to depend on the details of the
dynamics, i.e., it should be typical of systems with couplings
of the form $1/r^\alpha$, e.g. classical n-vector ferromagnets
(of which $n=2$ is the present case),
as long as the perturbation $\hat a$ has zero mean.
Systems for which the average perturbation is non-zero belong
to a different universality class, and alternative scaling laws
are expected \cite{b84,palv86,parv86,lima90}.


\section{Concluding remarks}


The scaling behavior of $\lambda_{max}$ with
$N \to \infty$ and $J \to 0$ [Eq.~(\ref{finalresult})] is exactly the same
as that obtained by using the geometric method \cite{fr01}.
The agreement can be also extended to the energy domain
($\varepsilon\rightarrow\infty$) by relating the time scale $\tau$ to
energy.
Given that the potential energy
is bounded, when $\varepsilon\rightarrow\infty$, the total energy and the
kinetic energy are essentially the same. In this regime, changing the time
scale is equivalent to a change in the kinetic energy, so that we have
$ \tau \propto \varepsilon^{-1/2}$. Thus we obtain the scaling law
$\lambda_{max} \propto \varepsilon^{-1/6}$.

The theoretical results (\ref{kappa}) agree with the
numerical ones obtained numerically in Refs.~\cite{at98,gmc00,cgmt00}.
There are some deviations which are consistent with finite size effects,
as argued in \cite{fr01}.
However, one should not discard the possibility that the scaling laws are
not exactly those derived in this paper (or in \cite{fr01}). 
The differences with numerical calculations might be due to the fact 
that both the geometric and the random matrix approaches assume ergodicity 
and fast (exponential) decay of the
correlations. We do not know at present if the dynamical system
fully satisfies these hypotheses. Eventually this issue will be decided when
simulations on larger systems are available.


\section*{Acknowledgements}
We thank S. Ruffo and C. Tsallis for useful comments.
We are grateful to M.-C. Firpo and S. Ruffo for communicating their
results prior to publication.
We acknowledge Brazilian Agencies CNPq, FAPERJ and PRONEX for financial
support.


\end{multicols}


\begin{thebibliography}{99}

\bibitem{ator98} M. Antoni and A. Torcini, Phys. Rev. E {\bf 57}, R6233
                 (1998).

\bibitem{tora99} A. Torcini and M. Antoni, Phys. Rev. E {\bf 59}, 2746
                 (1999).

\bibitem{lrr99}  V. Latora, A. Rapisarda, and S. Ruffo, Phys. Rev. Lett.
                 {\bf 83}, 2104 (1999).

\bibitem{lrr00}  V. Latora, A. Rapisarda, and S. Ruffo, Prog. Theo. Phys.
                 Supp. {\bf 139}, 204 (2000).

\bibitem{lrt00}  V. Latora, A. Rapisarda, and C. Tsallis, preprint
                 cond-mat/00103540.

\bibitem{ct01}   C. Tsallis, in "Nonextensive Statistical Mechanics and its
                 Applications", Series Lecture Notes in Physics
                 (Springer-Verlag,
                 Heidelberg, 2001).
                 See http://tsallis.cat.cbpf.br/biblio.htm for further
                 bibliography on the subject.

\bibitem{ccp99}  L. Casetti, E. G. D. Cohen, and M. Pettini, Phys. Rev.
                 Lett. {\bf 82} 4160 (1999).

\bibitem{at98}   C. Anteneodo and C. Tsallis, Phys. Rev. Lett. {\bf 80},
                 5313 (1998).

\bibitem{gmc00}  A. Giansanti, D. Moroni, and A. Campa, preprint
                 cond-mat/0007422.

\bibitem{cgmt00} A. Campa, A. Giansanti, D. Moroni, and C. Tsallis,
                 Phys. Lett. A {\bf 286}, 251 (2001).

\bibitem{ct95}   C. Tsallis, Fractals {\bf 3}, 541 (1995).

\bibitem{ar95}   M. Antoni and S. Ruffo, Phys. Rev. E {\bf 52}, 2361 (1995).

\bibitem{lrr98}  V. Latora, A. Rapisarda, and S. Ruffo, Phys. Rev. Lett.
                 {\bf 80}, 692 (1998).

\bibitem{eklr94} D. Escande, H. Kantz, R. Livi, and S. Ruffo,
                 J. Stat. Phys. {\bf 76}, 605 (1994).

\bibitem{ta00}   F. Tamarit and C. Anteneodo, Phys. Rev. Lett.
                 {\bf 84}, 208 (2000).

\bibitem{cgm00}  A. Campa, A. Giansanti, and D. Moroni, Phys. Rev. E
                 {\bf 62}, 303 (2000).

\bibitem{cpc00}  L. Casetti, M. Pettini, and E. G. D. Cohen, Phys. Rep.
                 {\bf 337}, 238 (2000).

\bibitem{ccp96}  L. Casetti, C. Clementi, and M. Pettini, Phys. Rev.
                 E {\bf 54} 5969 (1996).

\bibitem{f98}    M.-C. Firpo, Phys. Rev. E {\bf 57}, 6599 (1998).

\bibitem{fr01}   M.-C. Firpo and S. Ruffo, preprint cond-mat/0108158.

\bibitem{b84}    G. Benettin, Physica D {\bf 13}, 211 (1984).

\bibitem{palv86} G. Paladin and A. Vulpiani, J. Phys. A {\bf 19}, 1881 (1986).

\bibitem{parv86} G. Parisi and A. Vulpiani, J. Phys. A {\bf 19}, L425 (1986).

\bibitem{lima90} R. Lima and M. Rahibe, J. Phys. A: Math. Gen. {\bf 23}, 781 (1990).

\bibitem{livi}   R. Livi, A. Politi, S. Ruffo, and A. Vulpiani, J. Stat.
                 Phys. {\bf 46}, 147 (1987).

\end{thebibliography}
\end{document}